# NODE ISOLATION PROBABILITY OF WIRELESS ADHOC NETWORKS IN NAGAKAMI FADING CHANNEL

A.V. Babu[1] and Mukesh Kumar Singh[2]

[1,2]Department of Electronics & Communication Engineering,

National Institute of Technology Calicut, India

[1] babu@nitc.ac.in, [2] mukesh@gmail.com

## ABSTRACT

*This paper investigates the issue of connectivity of a wireless adhoc network in the presence of channel impairments. We derive analytical expressions for the node isolation probability in an adhoc network in the presence of Nakagami-m fading with superimposed lognormal shadowing. The node isolation probability is the probability that a randomly chosen node is not able to communicate with none of the other nodes in the network. An extensive investigation into the impact of path loss exponent, lognormal shadowing, Nakagami fading severity index, node density, and diversity order on the node isolation probability is conducted. The presented results are beneficial for the practical design of ad hoc networks.*

.

## 1. INTRODUCTION

In self-organizing wireless multihop ad-hoc networks such as sensor networks [1], the mobile devices communicate with each other in a peer-to-peer fashion without the need for base stations. For such networks, the level of connectivity among the mobile nodes depends on their spatial density, transmission and reception capabilities, and characteristics of the wireless channel. To achieve a fully connected adhoc network, there must be a path from any node to any other node. This paper analyze the connectivity of multihop radio networks in the presence of Nakagami fading with superimposed lognormal shadowing by computing a metric called node isolation probability. The node isolation probability is the probability that a randomly chosen node is not able to communicate with none of the other nodes in the network. The network becomes fully connected if there are no isolated nodes [8].

One of the first papers to address the connectivity issues in multihop networks was [2] in which authors investigated how far a node's broadcast message percolates, if the nodes are randomly distributed according to a homogeneous Poisson point process on an infinitely large area. The connectivity issues for nodes that are randomly distributed according to a uniform probability distribution on a one-dimensional line segment were addressed in [3]. Gupta and Kumar [4] performed a fundamental study on the connectivity of uniformly distributed nodes on a circular area. Penrose [5] also proved similar results independently. Santi and Blough [6, 7] conducted analytical investigations of the connectivity in bounded areas. The critical transmitting range for connectivity in an adhoc network in the presence of node mobility was first addressed in [7]. Bettsetter and Hartmann [8] addressed the impact of lognormal shadowing on the connectivity of adhoc networks. Work in [9] also addressed the same issue independently. Orriss and Barton [10, 11] obtained the connectivity results for the case of superposition of shadowing and fading phenomena. Haenggi [12] studied the impact of Rayleigh fading on network connectivity. The impact of interference on connectivity was analyzed in [13]. Miorandi *et al*. [14] presented analytical solution for network connectivity in the presence of channel randomness. Authors of [15] addressed the problem of finding the critical density of sensors required to achieve complete coverage of a desired region. Xiaole Bai *et. al*. [21] addressed the problem of determining an optimal deployment pattern that achieves both coverage and $k$-connectivity in a wireless sensor network. In [22], authors investigate the connectivity problem when directional antennas are used. While authors of [23] consider how physical layer cooperation can be used to improve the connectivity in wireless ad hoc networks.

In this paper, we derive analytical expression for node isolation probability in the presence of Nakagami fading and lognormal shadowing. We also analyze the impact of diversity combining schemes as well. Both maximal ratio combining (MRC) and selection combining (SC) schemes are considered for the analysis. The remainder of the paper is organized as follows. In Section 2, the preliminary assumptions and model are provided. Analytical evaluation of node isolation probability is presented in Section 3. Section 4 describes the numerical and simulation results. The paper is concluded in Section 5.

## 2. SYSTEM MODEL

We assume that the nodes in the network are randomly distributed according to a homogeneous Poisson point process and let $\lambda$ be the expected number of nodes per unit square ($0 < \lambda < \infty$). Let N be two-dimensional stationary Poisson point process over $\Re^2$. The points of the process represent the location of the nodes. Given a finite subset $A \in \Re^2$ of size $\upsilon(A)$, the number of nodes in A denoted by N(A) is a Poisson random variable with intensity $\lambda \upsilon(A)$. The numbers of nodes in disjoint areas are independent random variables. As in [8], we also neglect the impact of interference from other nodes. Let $P_I$ be the node isolation probability. The radio link is assumed Boolean: i.e., two nodes are either perfectly connected, or out of range. A switched link model is based on the assumption that the transmission between two nodes $l$ and $l'$ succeeds if and only if the signal to noise ratio (SNR) $\gamma$ at the receiver is greater than the threshold value $\psi$. For a received average SNR $y$, let $P_S(y)$ be the probability that the received instantaneous SNR $\gamma$ is greater than the threshold $\psi$. If good long codes are used, the function $P_S(y)$ approaches a step function [16]. Further let $R$ be the communication range of a node. In the presence of lognormal shadowing and small scale fading, $R$ is a random variable with cumulative distribution function (CDF) $F_R(\rho)$ and second moment $E[R^2]$. Since $R$ is non-negative, $E[R^2] = \int_0^\infty 2\rho \, d\rho \, F_R^c(\rho)$ where $F_R^c(\rho)$ is the complimentary CDF. The node isolation probability is then given by [8, 14]

$$P_I = e^{-\lambda \pi E[R^2]} \tag{1}$$

### 2.1. Combined Path-loss and Lognormal Shadowing

Assume that all the nodes transmit at a fixed power level $P_{tx}$ and let $W$ be the total white noise power present at the receiver. When lognormal shadowing is present, the mean of path loss is described by $K\rho^{-\alpha}$ where $\rho$ is the transmitter-receiver separation; $\alpha$-path loss exponent; and $K$ is a constant. The received SNR for this channel model is given by $\gamma(\rho) = \left( \dfrac{P_{tx}l(\rho)}{W} \right)$ where $l(\rho)$ is the path-loss. The communication range $R$ is the distance at which the SNR falls below the threshold $\psi$. Now $F_R(\rho)$ and $E[R^2]$ are computed as [14]

$$F_R(\rho) = P[\gamma(\rho) \leq \psi] = P[l(\rho) \leq \frac{W\psi}{P_{tx}}] = 1 - \int\limits_{\frac{W\psi}{P_{tx}}}^{\infty} f_{l/r}(a/\rho)da \qquad (2)$$

$$E[R^2] = \int\limits_{0}^{\infty} 2\rho \, d\rho \int\limits_{\frac{\psi W}{P_{tx}}}^{\infty} f_{l/r}(a/\rho)da \qquad (3)$$

where $f_{l/r}(a/\rho)$ represents the probability density function (PDF) of path loss under lognormal shadowing (with standard deviation $\sigma$) and is given by:

$$f_{l/r}(a/\rho) = \frac{1}{\sqrt{2\pi}\sigma a} e^{-\frac{1}{2}\left(\frac{\ln a - \ln(K\rho^{-\alpha})}{\sigma}\right)^2} \qquad (4)$$

## 2.2. Small Scale Fading and Lognormal Shadowing

Considering the effect of small scale fading alone, let $\gamma$ be the received instantaneous SNR and $y = E[\gamma]$ be the average SNR. For received average SNR $y = \left(\frac{KP_{tx}\rho^{-\alpha}}{W}\right)$, let $P_S(y)$ be the probability that the received instantaneous SNR $\gamma$ with PDF $f_\gamma(x/y)$ is greater than $\psi$. Now $P_S(y)$ and $E[R^2]$ are computed as

$$P_S(y) = \int\limits_{\psi}^{\infty} f_\gamma(x/y)\,dx \qquad (5)$$

$$E[R^2] = \int\limits_{0}^{\infty} 2\rho \, d\rho \, P_S\left(\frac{KP_{tx}\rho^{-\alpha}}{W}\right) \qquad (6)$$

For small scale fading with lognormal shadowing, $F_R(\rho)$ and $E[R^2]$ are evaluated as

$$F_R(\rho) = 1 - \int\limits_{0}^{\infty} P_S(y) f_{l/r}(a/\rho)da \qquad (7)$$

$$E[R^2] = \int\limits_{0}^{\infty} da \int\limits_{0}^{\infty} 2\rho \, d\rho \, P_S\left(\frac{aP_{tx}}{W}\right) f_{l/r}(a/\rho) \qquad (8)$$

## 3. NODE ISOLATION PROBABILITY ANALYSIS

In this section, analytical expressions are derived for node isolation probability of a wireless adhoc network in the presence of Nakagami fading with lognormal shadowing.

### 3.1. Nakagami Fading Channel

The PDF of the received signal envelope $Z$ under Nakagami-$m$ fading is [17]

$$f_Z(z) = \frac{2m^m z^{2m-1}}{\Gamma(m) y^m} e^{-\frac{mz^2}{y}}; m \geq 0.5; z \geq 0 \tag{9}$$

where $y$ is the average received power and $\Gamma(.)$ is the Gamma function [18]. The PDF of received instantaneous SNR $\gamma$ is [17]

$$f_\gamma(x/y) = \frac{m^m x^{m-1}}{y^m \Gamma(m)} e^{-\frac{mx}{y}}; m \geq 0.5; x \geq 0 \tag{10}$$

For Nakagami fading, the success probability $P_S(y)$ is computed as

$$P_S(y) = \int_\psi^\infty \frac{m^m x^{m-1}}{y^m \Gamma(m)} e^{-\frac{mx}{y}} dx = \frac{\Gamma(m, m\psi/y)}{\Gamma(m)} \tag{11}$$

where $\Gamma(m, m\psi/y) = \int_{m\psi/y}^\infty t^{m-1} e^{-t} dt$ is the incomplete gamma function. Assuming $m$ to take positive integer values, the success probability $P_S(y)$ becomes

$$P_S(y) = e^{-m\psi/y} \sum_{l=0}^{m-1} \frac{1}{l!} \left( \frac{m\psi}{y} \right)^l \tag{12}$$

In the absence of lognormal shadowing, $E[R^2]$ is computed by substituting (12) in (6) and is given by:

$$E[R^2] = \int_0^\infty 2\rho \, d\rho \, P_S \left( \frac{KP_{tx} \rho^{-\alpha}}{W} \right) = \int_0^\infty 2\rho \, d\rho \, e^{-\frac{m\psi W \rho^\alpha}{KP_{tx}}} \sum_{l=0}^{m-1} \frac{\rho^{\alpha l}}{l!} \frac{(m\psi W)^l}{(KP_{tx})^l}$$

$$= \sum_{l=0}^{m-1} \frac{(m\psi)^l}{l!} \left( \frac{W}{KP_{tx}} \right)^l \int_0^\infty 2\rho^{\alpha l+1} e^{-\frac{m\psi W}{KP_{tx}} \rho^\alpha} d\rho \tag{13}$$

Now (13) is simplified by using the following result (14) from [18] and the simplified expression is given in (15).

$$\int_0^\infty x^{\upsilon-1} e^{-\mu x^p} dx = \frac{\mu^{-\upsilon/p}}{|p|} \Gamma\left(\frac{\upsilon}{p}\right) \tag{14}$$

$$E[R^2] = \frac{2}{\alpha} \left(\frac{m\psi W}{KP_{tx}}\right)^{-2/\alpha} \sum_{l=0}^{m-1} \frac{1}{l!} \Gamma\left(\frac{2}{\alpha}+l\right) \tag{15}$$

The node isolation probability is then determined by combining (1) and (15) and is given by

$$P_I = \exp\left\{ -\lambda\pi \frac{2}{\alpha} \left(\frac{m\psi W}{KP_{tx}}\right)^{-2/\alpha} \sum_{l=0}^{m-1} \frac{1}{l!} \Gamma\left(\frac{2}{\alpha}+l\right) \right\} \tag{16}$$

### 3.2 Nakagami Fading with Superimposed Lognormal Shadowing

Next we consider the impact of Nakagami fading with lognormal shadowing. Given that $y$ is the received average SNR, $E[R^2]$ is computed by substituting (4) and (12) in (8).

$$\begin{aligned} E[R^2] &= \int_0^\infty da \int_0^\infty 2\rho\, d\rho\, P_S\left(y = \frac{aP_{tx}}{W}\right) f_{l/r}(a/\rho) \\ &= \int_0^\infty da \int_0^\infty 2\rho\, d\rho\, e^{-\frac{m\psi W}{aP_{tx}}} \sum_{l=0}^{m-1} \frac{(m\psi)^l}{l!} \left(\frac{W}{aP_{tx}}\right)^l \frac{1}{\sqrt{2\pi}\sigma a} e^{-\frac{1}{2}\left(\frac{\ln a - \ln(K\rho^{-\alpha})}{\sigma}\right)^2} \end{aligned} \tag{17}$$

Let $x = \frac{1}{\sigma} \ln(a\rho^\alpha / K)$, then the above integral is simplified as

$$E[R^2] = \int_{-\infty}^\infty dx \frac{1}{\sqrt{2\pi}} e^{\frac{-x^2}{2}} \int_0^\infty 2\rho\, d\rho\, e^{-\frac{m\psi W e^{-x\sigma}\rho^\alpha}{KP_{tx}}} \sum_{l=0}^{m-1} \frac{\rho^{\alpha l}}{l!} \left(\frac{m\psi W e^{-x\sigma}}{KP_{tx}}\right)^l \tag{18}$$

Now (18) is simplified using (14) and is given by

$$\begin{aligned} i.e.;\ E[R^2] &= \int_{-\infty}^\infty dx \frac{1}{\sqrt{2\pi}} e^{-\frac{x^2}{2}} \frac{2}{\alpha} \left(\frac{m\psi W e^{-x\sigma}}{KP_{tx}}\right)^{-2/\alpha} \sum_{l=0}^{m-1} \frac{1}{l!} \Gamma\left(\frac{2}{\alpha}+l\right) \\ &= \frac{2}{\alpha} \left(\frac{m\psi W}{KP_{tx}}\right)^{-2/\alpha} e^{\frac{2\sigma^2}{\alpha^2}} \sum_{l=0}^{m-1} \frac{1}{l!} \Gamma\left(\frac{2}{\alpha}+l\right) \end{aligned} \tag{19}$$

The node isolation probability is obtained by combining (1) and (19):

$$P_I = \exp\left\{ -\lambda\pi \frac{2}{\alpha} \left(\frac{m\psi W}{KP_{tx}}\right)^{-2/\alpha} e^{\frac{2\sigma^2}{\alpha^2}} \sum_{l=0}^{m-1} \frac{1}{l!} \Gamma\left(\frac{2}{\alpha}+l\right) \right\} \tag{20}$$

### 3.3. MRC with Independent Nakagami Fading

In MRC, the $M$ received signals are combined such that the output SNR is maximized. We assume equal $m$ for all the diversity branches and identical average SNR on each branch equal to $y$. The instantaneous SNR at the output of MRC is $\gamma = \sum_{k=1}^{M} \gamma_k$ where $\gamma_k$ is the SNR in branch $k$. With statistically independent Nakagami faded branches, the PDF of $\gamma$ is given by [19]

$$f_\gamma \left( x/y \right) = \left( \frac{m}{y} \right)^{mM} \frac{x^{mM-1}}{\Gamma(mM)} e^{-\frac{mx}{y}}; m \geq 0.5; x \geq 0 \tag{21}$$

Combining (5) and (21) and assuming $m$ to take positive integer values, $P_S(y)$ is obtained as

$$P_S \left( y \right) = \frac{\Gamma(mM, m\psi/y)}{\Gamma(mM)} = e^{-m\psi/y} \sum_{l=0}^{mM-1} \frac{1}{l!} \left( \frac{m\psi}{y} \right)^l \tag{22}$$

Now $E[R^2]$ is obtained by substituting (22) in (8). Following the procedure adopted for the derivation of (19) and (20), $E[R^2]$ and $P_I$ are obtained as

$$E[R^2] = \frac{2}{\alpha} \left( \frac{m\psi W}{KP_{tx}} \right)^{-2/\alpha} \exp \left( \frac{2\sigma^2}{\alpha^2} \right) \sum_{l=0}^{mM-1} \frac{1}{l!} \Gamma \left( \frac{2}{\alpha} + l \right) \tag{23}$$

$$P_I = \exp \left\{ -\lambda \pi \frac{2}{\alpha} \left( \frac{m\psi W}{KP_{tx}} \right)^{-2/\alpha} e^{\frac{2\sigma^2}{\alpha^2}} \sum_{l=0}^{mM-1} \frac{1}{l!} \Gamma \left( \frac{2}{\alpha} + l \right) \right\} \tag{24}$$

### 3.4. SC with Independent Nakagami Fading

In SC, the combiner chooses and processes only the branch with the highest SNR. The combined branches are assumed to be independent of each other and have the same average SNR. Given that $\gamma_k$ is the SNR in branch $k$, the instantaneous SNR at the output of SC is given by $\gamma_{sc} = \max \left( \gamma_1, \gamma_2, ..., \gamma_M \right)$. The CDF of $\gamma_{sc}$ is [20]

$$F_\gamma(x/y) = \left[1 - \exp\left(\frac{-mx}{y}\right)\sum_{k=0}^{m-1}\left(\frac{mx}{y}\right)^k \frac{1}{k!}\right]^M$$

$$= \sum_{n=0}^{M}(-1)^n M_{C_n}\exp\left(\frac{-nmx}{y}\right)\left[\sum_{k=0}^{m-1}\left(\frac{mx}{y}\right)^k \frac{1}{k!}\right]^n$$

$$= \sum_{n=0}^{M}(-1)^n M_{\text{£}_n}\exp\left(\frac{-nmx}{y}\right)\sum_{k=0}^{n(m-1)}\beta_{kn}\left(\frac{mx}{y}\right)^k \tag{25}$$

The final expression in (25) is obtained after expanding the first expression in (25) binomially and then using multinomial theorem. Here $\beta_{kn}$ is determined as

$$\beta_{kn} = \sum_{i=k-m+1}^{k}\frac{\beta_{i(n-1)}}{(k-i)!}I_{[0,(n-1)(m-1)]}(n); \beta_{00} = \beta_{0n} = 1, \beta_{k1} = \frac{1}{k!}$$

$$I_{[a,b]}(n) = \begin{cases} 1, a \le n \le b \\ 0, otherwise \end{cases} \tag{26}$$

The success probability, which is the probability that SNR in at least one path is greater than $\psi$, is the complement of the probability of all paths presenting an SNR lower than $\psi$ and is given by

$$P_S(y) = -\sum_{n=1}^{M}(-1)^n M_{\text{£}_n}\exp\left(\frac{-nm\psi}{y}\right)\sum_{k=0}^{n(m-1)}\beta_{kn}\left(\frac{m\psi}{y}\right)^k \tag{27}$$

Expression for $E[R^2]$ is obtained by substituting (4) and (27) in (8). Final expressions for $E\left[R^2\right]$ and $P_I$ are as follows:

$$E\left[R^2\right] = \left\{-\frac{2}{\alpha}\left(\frac{m\psi W}{KP_{tx}}\right)^{\frac{-2}{\alpha}}e^{\frac{2\sigma^2}{\alpha^2}}\sum_{h=1}^{M}(-1)^h M_{\text{£}_h}\sum_{l=0}^{h(m-1)}\beta_{lh}h^{-\left(\left(\frac{2}{\alpha}\right)+l\right)}\Gamma\left(\left(\frac{2}{\alpha}\right)+l\right)\right\} \tag{28}$$

$$P_I = \exp\left\{\lambda\pi\frac{2}{\alpha}\left(\frac{m\psi W}{KP_{tx}}\right)^{\frac{-2}{\alpha}}e^{\frac{2\sigma^2}{\alpha^2}}\sum_{h=1}^{M}(-1)^h M_{\text{£}_h}\sum_{l=0}^{h(m-1)}\beta_{lh}h^{-\left(\left(\frac{2}{\alpha}\right)+l\right)}\Gamma\left(\left(\frac{2}{\alpha}\right)+l\right)\right\} \tag{29}$$

## 4. NUMERICAL AND SIMULATION RESULTS

In this section we present the numerical and simulation results. The numerical results are obtained from the analytical model using MATLAB. The system parameters are selected as follows: $K$=10dB, $P_{tx}$=1mWatt, $W$=0.01mWatt, $\psi$=10dB. The parameters such as $m, \lambda, \alpha$, and $\sigma$ are selected suitably. Simulation is also performed using MATLAB for the same system parameters. The network size for simulation is fixed to be $100m \times 100m$. We choose a random number of nodes according to Poisson process and the nodes are placed over the simulation area according to a random uniform distribution. Various links are then established according to the given channel model. The nodes operate in a Nakagami fading environment with super imposed lognormal shadowing. In this topology, we check for an isolated node and the experiment is repeated for many random topologies. The node isolation probability is computed as an average of 1000 simulation runs.

The simulated network topologies are shown in Figures 1(a) – 1(d) for various channel conditions. Figures 2 and 3 show the node isolation probability $P_I$ versus node density $\lambda$. In Figure 2, $\sigma$ is chosen as the variable while Figure 3 is drawn for different values of $m$. Figure 4 shows the relation between $\lambda$ and $\sigma$ for a fixed $P_I$. Larger values of $\lambda$ result in lower $P_I$. Further, for larger values of $\sigma$, a lower $\lambda$ is sufficient to make the network connected with the same probability. In other words, assuming the path loss exponent $\alpha$ to be fixed, larger values of $\sigma$ always makes the network to become connected, without any increase in transmitted power. For $m = 1$, the results correspond to Rayleigh fading and for higher values of $m$, $P_I$ decreases. It may be noted that the difference between analysis and simulation results are marginal. Figure 5 shows impact of $\alpha$ on $P_I$, keeping all other parameters to be constant. For a fixed value of lognormal spread $\sigma$, higher value of path loss exponent $\alpha$ always results in larger isolation probability. General multi-path fading always correspond to $\alpha > 2$ ($\alpha = 2$ correspond to free-space propagation). Hence we conclude that larger values of lognormal spread reduce the node isolation probability. Further, as the Nagakami fading factor increases, the node isolation probability gets reduced, improving the connectivity of the network.

Next, the performance evaluation is repeated for receive diversity with MRC and SC schemes. The results for independent MRC scheme are shown in Figure 6, while Figure 7 shows the results for independent SC scheme. Both MRC and SC schemes improve the network connectivity performance. For the MRC scheme, the percentage improvement in $P_I$ increases as $M$ increases, while for SC scheme; it is found that there is no proportional improvement in $P_I$ for higher values of $M$. Thus the presented model can be used to find, for a given set of channel model parameters, the minimum node density required to achieve a fully connected network covering a certain area.

## 5. SUMMARY

In this paper, we have investigated connectivity properties of multi-hop wireless networks by computing a metric called node isolation probability. Analytical expressions for node isolation probability were derived. As opposed to previous research in this field, we took into account stochastic lognormal shadowing, and Nakagami fading effects between the nodes. Using a combination of analytical and simulation-based methods, we gave insight about the impact of various parameters such as node density, path loss exponent, lognormal spread, and Nakagami-$m$ factor on the isolation probability. We have also investigated the effect of diversity-combining techniques on the network connectivity performance. The computed values of the node density are of practical relevance for the design and simulation of wireless multi-hop networks. For a given channel model and parameters, the presented results can be used to determine the minimum node density that is needed to achieve a fully connected network covering a certain area.

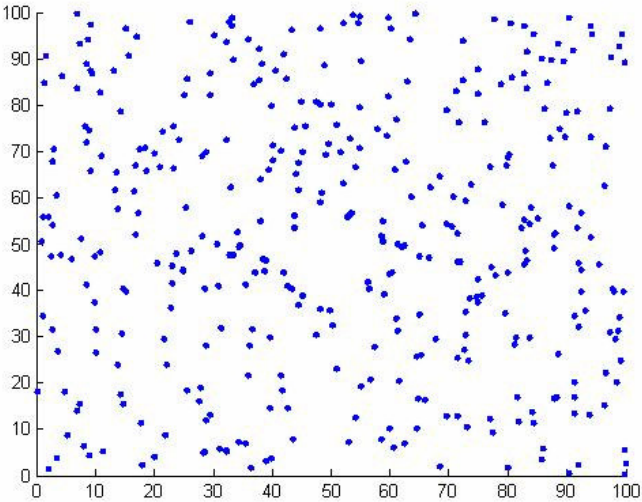

Figure 1(a). Simulated Network (Nakagami fading, $\sigma = 0$)

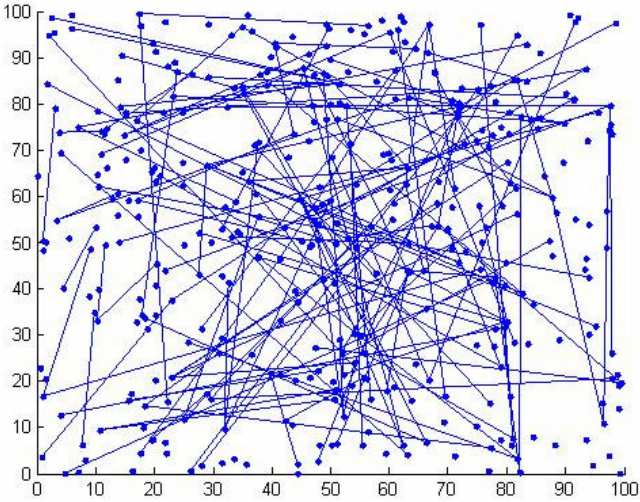

Figure 1(b). Simulated Network (Nakagami fading, $\sigma = 2$)

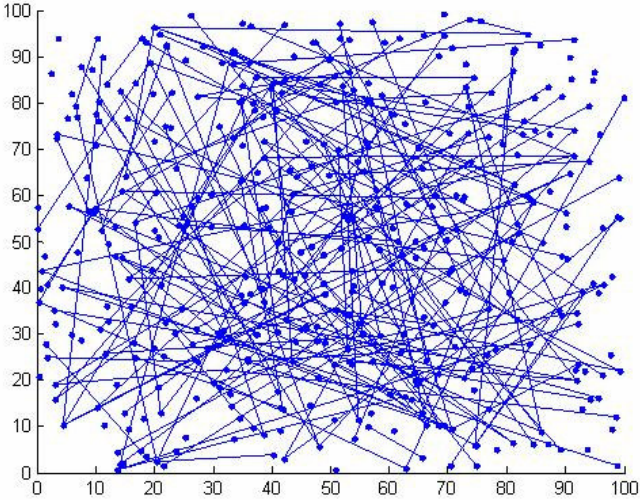

Figure 1(c). Simulated Network (Nakagami fading, $\sigma = 4$)

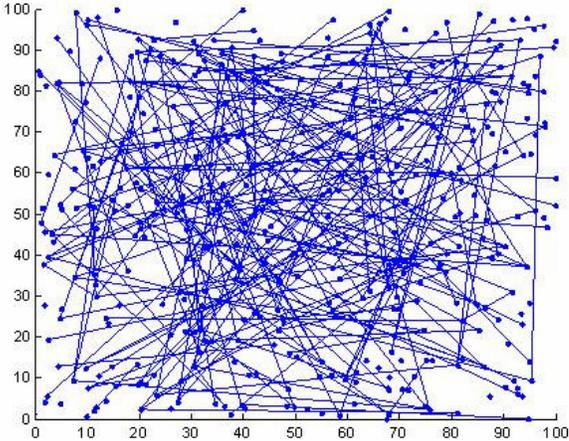

Figure 1(d). Simulated Network (Nakagami fading with MRC, $\sigma = 2$, $M = 2$)

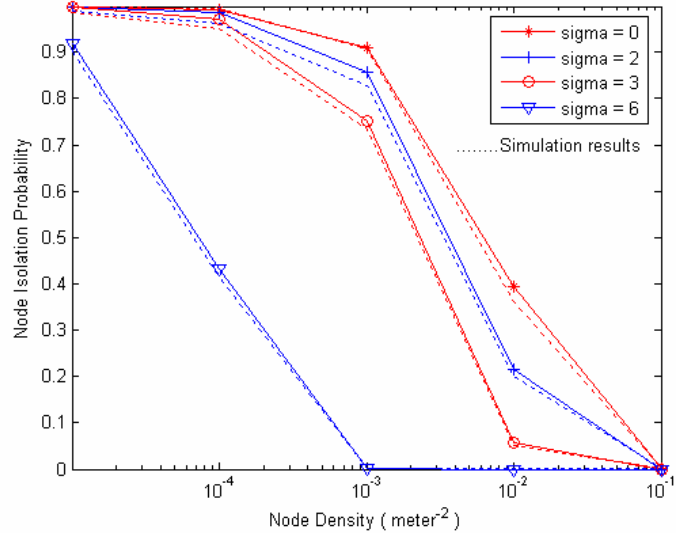

Figure 2. Node isolation probability $P_I$ vs. node density $\lambda$
(Nakagami fading $m$=2, $\alpha$=4

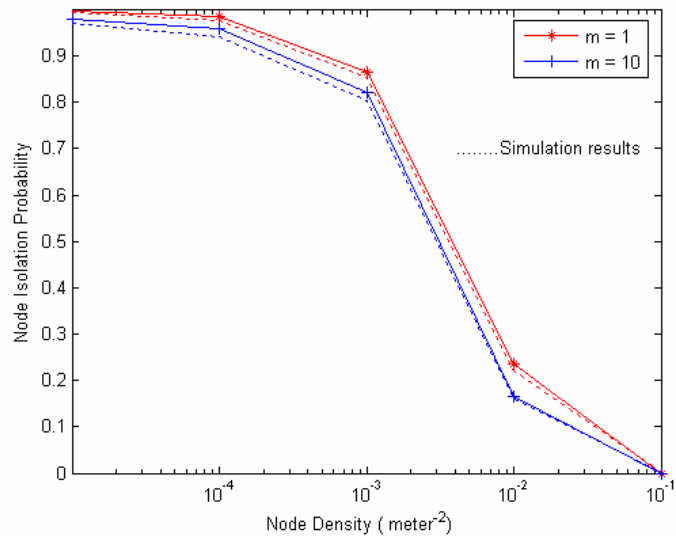

Figure 3. Node isolation probability $P_I$ vs. node density $\lambda$
(Nakagami fading $\sigma$=2, $\alpha$=4)

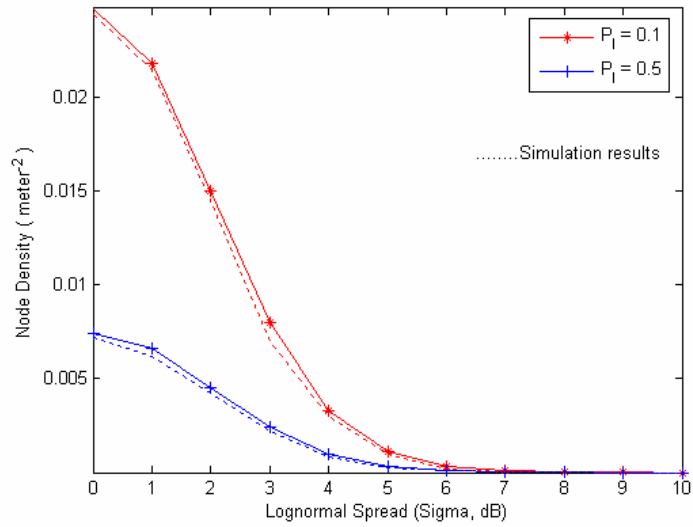

Figure 4. Node density $\lambda$ vs. lognormal spread $\sigma$ : Nakgami fading ($m$=4, $\alpha$ =4)

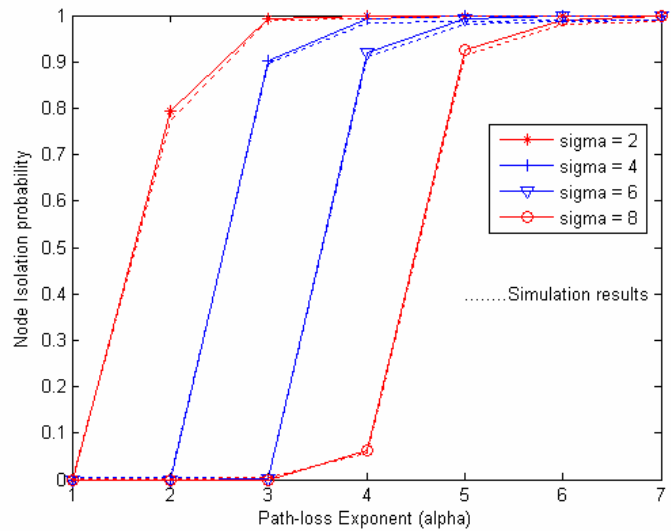

Figure 5. $P_I$ vs. path loss exponent $\alpha$ : Nakagami fading ($m = 4$, $\lambda$ =.00001 / $m^{-2}$ )

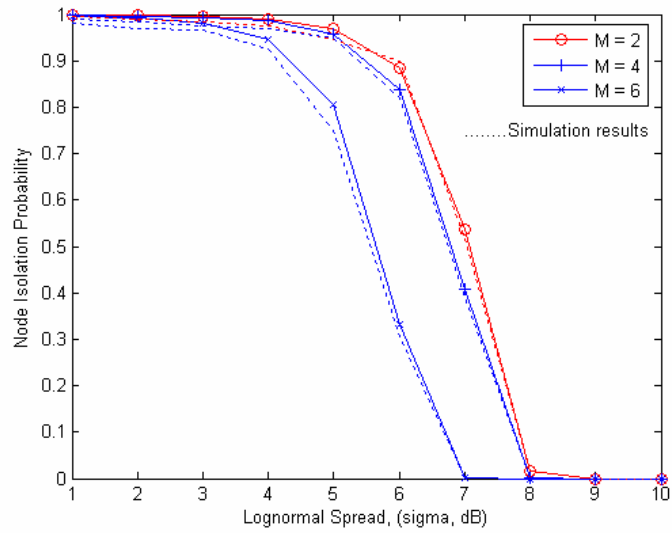

Figure 6. $P_I$ vs. lognormal spread (Receive diversity with MRC: $m=2$, $\lambda =.00001 / m^{-2}$, $\alpha =4$)

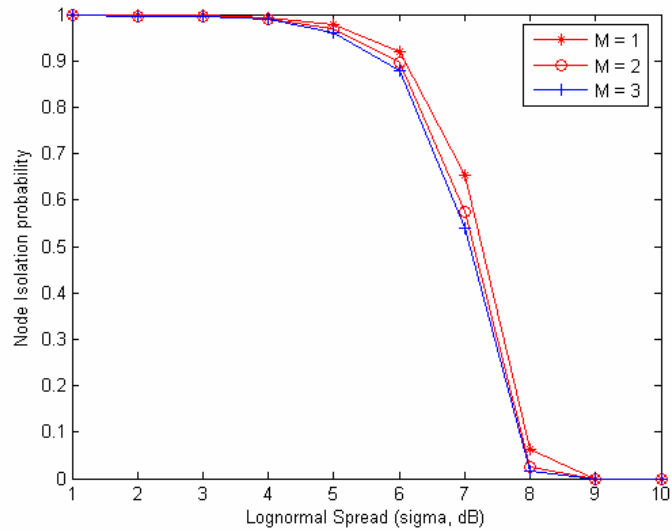

Figure 7. $P_I$ vs. lognormal spread (Receive diversity with SC: $m=2$, $\lambda =.00001 / m^{-2}$, $\alpha =4$)